# An Analysis of the Skype Peer-to-Peer Internet Telephony Protocol

Salman A. Baset and Henning Schulzrinne
Department of Computer Science
Columbia University, New York NY 10027
{salman,hgs}@cs.columbia.edu

September 15, 2004

## ABSTRACT
Skype is a peer-to-peer VoIP client developed by KaZaa in 2003. Skype claims that it can work almost seamlessly across NATs and firewalls and has better voice quality than the MSN and Yahoo IM applications. It encrypts calls end-to-end, and stores user information in a decentralized fashion. Skype also supports instant messaging and conferencing.

This report analyzes key Skype functions such as login, NAT and firewall traversal, call establishment, media transfer, codecs, and conferencing under three different network setups. Analysis is performed by careful study of Skype network traffic.

## Categories and Subject Descriptors
C.2.2 [**Computer-Communication Networks**]: Network Protocols—*Applications*

## General Terms
Algorithms, Design, Measurement, Performance, Experimentation, Security,

## Keywords
Peer-to-peer (p2p), Voice over IP (VoIP), Super Node (SN), Internet telephony, conferencing

## 1. INTRODUCTION
Skype is a peer-to-peer VoIP client developed by KaZaa [17] that allows its users to place voice calls and send text messages to other users of Skype clients. In essence, it is very similar to the MSN and Yahoo IM applications, as it has capabilities for voice-calls, instant messaging, audio conferencing, and buddy lists. However, the underlying protocols and techniques it employs are quite different.

Like its file sharing predecessor KaZaa, Skype is an overlay peer-to-peer network. There are two types of nodes in this overlay network, ordinary hosts and super nodes (SN). An ordinary host is a Skype application that can be used to place voice calls and send text messages. A super node is an ordinary host's end-point on the Skype network. Any node with a public IP address having sufficient CPU, memory, and network bandwidth is a candidate to become a super node. An ordinary host must connect to a super node and must register itself with the Skype login server for a successful login. Although not a Skype node itself, the Skype login server is an important entity in the Skype network. User names and passwords are stored at the login server. User authentication at login is also done at this server. This server also ensures that Skype login names are unique across the Skype name space. Figure 1 illustrates the relationship between ordinary hosts, super nodes and login server.

Apart from the login server, there is no central server in the Skype network. Online and offline user information is stored and propagated in a decentralized fashion and so are the user search queries.

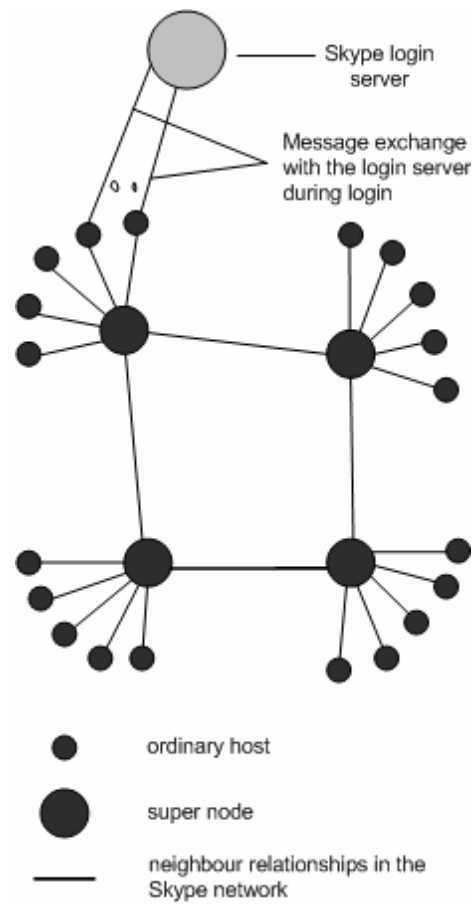

**Figure 1. Skype Network. There are three main entities: supernodes, ordinary nodes, and the login server.**

NAT and firewall traversal are important Skype functions. We believe that each Skype node uses a variant of STUN [1] protocol to determine the type of NAT and firewall it is behind. We also believe that there is no global NAT and firewall traversal server because if there was one, the Skype node would have exchanged traffic with it during login and call establishment in the many experiments we performed.

The Skype network is an overlay network and thus each Skype client (SC) should build and refresh a table of reachable nodes. In Skype, this table is called host cache (HC) and it contains IP address and port number of super nodes. It is stored in the Windows registry for each Skype node.

Skype claims to have implemented a '3G P2P' or 'Global Index' [2] technology (Section 4.3), which is guaranteed to find a user if that user has logged in the Skype network in the last 72 hours.

Skype uses wideband codecs which allows it to maintain reasonable call quality at an available bandwidth of 32 kb/s. It uses TCP for signaling, and both UDP and TCP for transporting media traffic. Signaling and media traffic are not sent on the same ports.

The rest of this report is organized as follows. Section 2 describes key components of the Skype software and the Skype network. Section 3 describes the experimental setup. Section 4 discusses key Skype functions like startup, login, user search, call establishment, media transfer and codecs, and presence timers. Flow diagrams based on actual network traffic are used to elaborate on the details. Section 5 discusses conferencing. Section 6 discusses other experiments.

## 2. KEY COMPONENTS OF THE SKYPE SOFTWARE

A Skype client listens on particular ports for incoming calls, maintains a table of other Skype nodes called host cache, uses wideband codecs, maintains a buddy list, encrypts messages end-to-end, and determines if it is behind a NAT or a firewall. This section discusses these components and functionalities in detail.

### 2.1 Ports

A Skype client (SC) opens a TCP and a UDP listening port at the port number configured in its connection dialog box. SC randomly chooses the port number upon installation. In addition, SC also opens TCP listening ports at port number 80 (HTTP port), and port number 443 (HTTPS port). Unlike many Internet protocols, like SIP [5] and HTTP [6], there is no default TCP or UDP listening port. Figure 15 shows a snapshot of the Skype connection dialog box. This figure shows the ports on which a SC listens for incoming connections.

### 2.2 Host Cache

The host cache (HC) is a list of super node IP address and port pairs that SC builds and refreshes regularly. It is the most critical part to the Skype operation. At least one valid entry must be present in the HC. A valid entry is an IP address and port number of an online Skype node. A SC stores host cache in the Windows registry at HKEY_CURRENT_USER / SOFTWARE / SKYPE / PHONE / LIB / CONNECTION / HOSTCACHE. After running a SC for two days, we observed that HC contained a maximum of 200 entries. Host and peer caches are not new to Skype. Chord [19], another peer-to-peer protocol has a finger table, which it uses to quickly find a node.

### 2.3 Codecs

The white paper [7] observes that Skype uses iLBC [8], iSAC [9], or a third unknown codec. GlobalIPSound [10] has implemented the iLBC and iSAC codecs and their website lists Skype as their partner. We believe that Skype uses their codec implementations. We measured that the Skype codecs allow frequencies between 50-8,000 Hz to pass through. This frequency range is the characteristic of a wideband codec.

### 2.4 Buddy List

Skype stores its buddy information in the Windows registry. Buddy list is digitally signed and encrypted. The buddy list is local to one machine and is not stored on a central server. If a user uses SC on a different machine to log onto the Skype network, that user has to reconstruct the buddy list.

### 2.5 Encryption

The Skype website [13] explains: "Skype uses AES (Advanced Encryption Standard) – also known as Rijndel – which is also used by U.S. Government organizations to protect sensitive information. Skype uses 256-bit encryption, which has a total of $1.1 \times 10^{77}$ possible keys, in order to actively encrypt the data in each Skype call or instant message. Skype uses 1536 to 2048 bit RSA to negotiate symmetric AES keys. User public keys are certified by Skype server at login."

### 2.6 NAT and Firewall

We conjecture that SC uses a variation of the STUN [1] and TURN [18] protocols to determine the type of NAT and firewall it is behind. We also conjecture that SC refreshes this information periodically. This information is also stored in the Windows registry.

Unlike its file sharing counter part KaZaa, a SC cannot prevent itself from becoming a super node.

## 3. EXPERIMENTAL SETUP

All experiments were performed for Skype version 0.97.0.6. Skype was installed on two Windows 2000 machines. One machine was a Pentium II 200MHz with 128 MB RAM, and the other machine was a Pentium Pro 200 MHz with 128 MB RAM. Each machine had a 10/100 Mb/s Ethernet card and was connected to a 100 Mb/s network.

We performed experiments under three different network setups. In the first setup, both Skype users were on machines with public IP addresses; in the second setup, one Skype user was behind port-restricted NAT; in the third setup, both Skype users were behind a port-restricted NAT and UDP-restricted firewall. NAT and firewall machines ran Red Hat Linux 8.0 and were connected to 100 Mb/s Ethernet network.

Ethereal [3] and NetPeeker [4] were used to monitor and control network traffic, respectively. NetPeeker was used to tune the bandwidth so as to analyze the Skype operation under network congestion.

For each experiment, the Windows registry was cleared of any Skype entries and Skype was reinstalled on the machine.

All experiments were performed between February and April, 2004.

## 4. SKYPE FUNCTIONS

Skype functions can be classified into startup, login, user search, call establishment and tear down, media transfer, and presence messages. This section discusses each of them in detail.

### 4.1 Startup

When SC was run for the first time after installation, it sent a HTTP 1.1 GET request to the Skype server (skype.com). The first line of this request contains the keyword 'installed'.

During subsequent startups, a SC only sent a HTTP 1.1 GET request to the Skype server (skype.com) to determine if a new version is available. The first line of this request contains the keyword 'getlatestversion'.

See the Appendix for complete messages.

### 4.2 Login

Login is perhaps the most critical function to the Skype operation. It is during this process a SC authenticates its user name and password with the login server, advertises its presence to other peers and its buddies, determines the type of NAT and firewall it is behind, and discovers online Skype nodes with public IP addresses. We observed that these newly discovered nodes were used to maintain connection with the Skype network should the SN to which SC was connected became unavailable.

#### 4.2.1 Login Process

As discussed in Section 2, the HC must contain a valid entry for a SC to be able to connect to the Skype network. If the HC was filled with only one invalid entry, SC could not connect to the Skype network and reported a login failure. However, we gained useful insights in the Skype login process by observing the message flow between SC and this invalid HC entry. The experimental setup and observations for the login process are described below.

First, we flushed the SC host cache and filled it with only one entry which was the IP address and port number of a machine on which no Skype client was running. The SC was then started and a login attempt was made. Since HC had an invalid entry, SC could not connect to the Skype network. We observed that the SC first sent a UDP packet to this entry. If there was no response after roughly five seconds, SC tried to establish a TCP connection with this entry. It then tried to establish a TCP connection to the HC IP address and port 80 (HTTP port). If still unsuccessful, it tried to connect to HC IP address and port 443 (HTTPS port). SC then waited for roughly 6 seconds. It repeated the whole process four more times after which it reported a login failure.

We observed that a SC must establish a TCP connection with a SN in order to connect to the Skype network. If it cannot connect to a super node, it will report a login failure.

Most firewalls are configured to allow outgoing TCP traffic to port 80 (HTTP port) and port 443 (HTTPS port). A SC behind a firewall, which blocks UDP traffic and permits selective TCP traffic, takes advantage of this fact. At login, it establishes a TCP connection with another Skype node with a public IP address and port 80 or port 443.

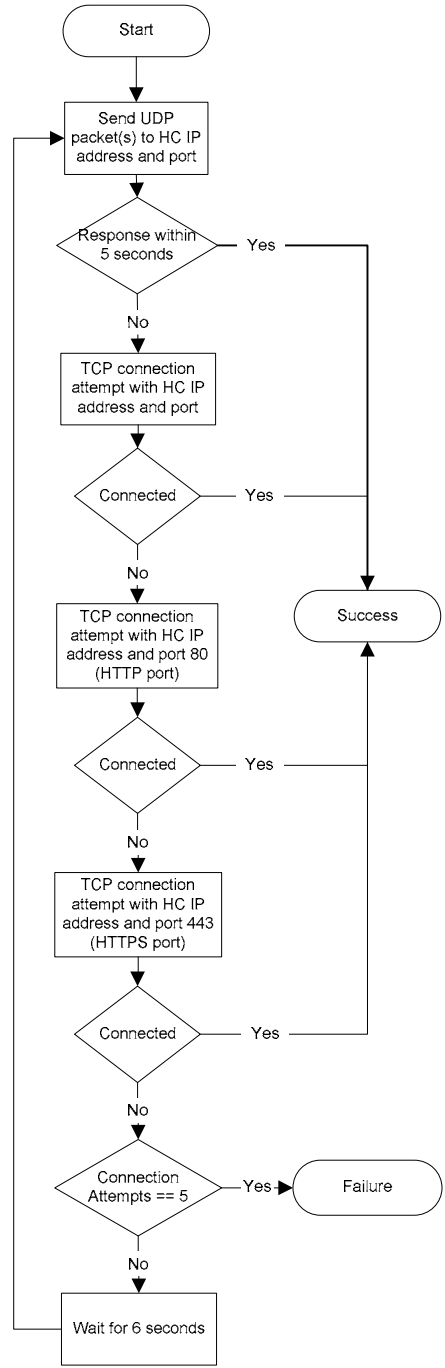

Figure 2. Skype login algorithm. Only one entry is present in the HC. If there is more than one entry, SC sends UDP packets to them before attempting a TCP connection. Authentication with the login server is not shown.

#### 4.2.2 Login Server

After a SC is connected to a SN, the SC must authenticate the user name and password with the Skype login server. The login server is the only central component in the Skype network. It stores Skype user names and passwords and ensures that Skype user names are unique across the Skype name space. SC must authenticate itself with login server for a successful login. During our experiments we observed that SC always exchanged data over TCP with a node whose IP address was 80.160.91.11. We believe that this node is the login server. A reverse lookup of this IP address retrieved NS records whose values are **ns14.inet.tele.dk** and **ns15.inet.tele.dk**. It thus appears from the reverse lookup that the login server is hosted by an ISP based in Denmark.

### 4.2.3 Bootstrap Super Nodes

After logging in for the first time after installation, HC was initialized with seven IP address and port pairs. We observed that upon first login, HC was always initialized with these seven IP address and port pairs except for a rare random occurrence. In the case where HC was initialized with more than seven IP addresses and port pairs, it always contained those seven IP address and port pairs. It was with one of these IP address and port entries a SC established a TCP connection when a user used that SC to log onto the Skype network for the first time after installation. We call these IP address and port pairs bootstrap super nodes. Figure 16 shows a snapshot of the host cache of the SC that contains IP address and port numbers of these bootstrap super nodes. These IP address and port pairs and their corresponding host names obtained using a reverse lookup are:

| IP address:port | Reverse lookup result |
|---|---|
| 66.235.180.9:33033 | sls-cb10p6.dca2.superb.net |
| 66.235.181.9:33033 | ip9.181.susc.suscom.net |
| 80.161.91.25:33033 | 0x50a15b19.boanxx15.adsl-dhcp.tele.dk |
| 80.160.91.12:33033 | 0x50a15b0c.albnxx9.adsl-dhcp.tele.dk |
| 64.246.49.60:33033 | rs-64-246-49-60.ev1.net |
| 64.246.49.61:33033 | rs-64-246-49-61.ev1.net |
| 64.246.48.23:33033 | ns2.ev1.net |

From the reverse lookup, it appears that bootstrap SNs are connected to the Internet through four ISPs. Superb [14], Suscom [15], ev1.net [16] are US-based ISPs.

After installation and first time startup, we observed that the HC was empty. However upon first login, the SC sent UDP packets to at least four nodes in the bootstrap node list. Thus, either bootstrap IP address and port information is hard coded in the SC, or it is encrypted and not directly visible in the Skype Windows registry, or this is a one-time process to contact bootstrap nodes. We also observed that if the HC was flushed after the first login, SC was unable to connect to the Skype network. These observations suggest that we perform separate experiments to analyze the first-time and subsequent login processes.

### 4.2.4 First-time Login Process

The SC host cache was empty upon installation. Thus, a SC must connect to well known Skype nodes in order to log on to the Skype network. It does so by sending UDP packets to some bootstrap super nodes and then waits for their response over UDP for some time. It is not clear how SC selects among bootstrap SNs to send UDP packets to. SC then established a TCP connection with the bootstrap super node that responded. Since more than one node could respond, a SC could establish a TCP connection with more than one bootstrap node. A SC, however, maintains a TCP connection with at least one bootstrap node and may close TCP connections with other nodes. After exchanging some packets with SN over TCP, it then perhaps acquired the address of the login server (80.160.91.11). SC then establishes a TCP connection with the login server, exchanges authentication information with it, and finally closes the TCP connection. The initial TCP data exchange with the bootstrap SN and the login server shows the existence of a challenge-response mechanism.

The TCP connection with the SN persisted as long as SN was alive. When the SN became unavailable, SC establishes a TCP connection with another SN.

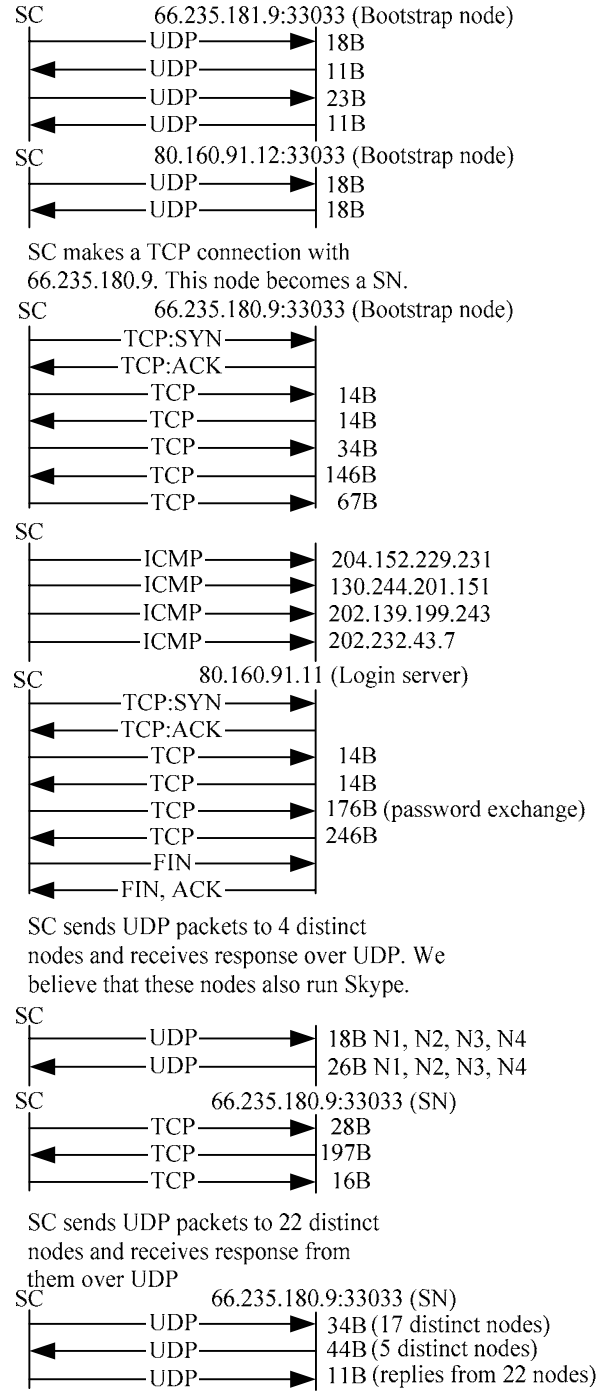

**Figure 3. Message flow for the first login after installation for SC on a public IP address. 'B' stands for bytes and 'N' stands

for node. SYN and ACK packets are shown to indicate who initiated TCP connection. Message flows are not strictly according to time. Messages have been grouped together to provide a better picture. Message size corresponds to size of TCP or UDP payload. Not all messages are shown.

For the login process, we observed message flow for the same Skype user id for the three different network setups described in Section 3.

The message flow for the first-time login process for a SC running on a machine with public IP address is shown in Figure 3. The total data exchanged between SC, SN, login server, and other nodes during login is about 9 kilobytes.

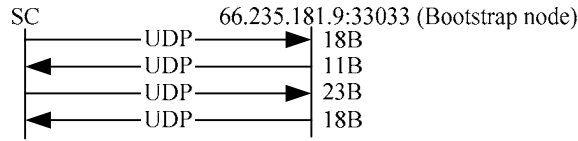

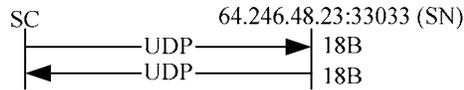

SC makes a TCP connection with 64.246.48.23. This node becomes a SN.

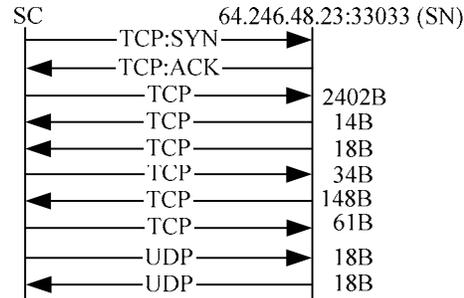

For same Skype user id, SC on public IP address and SC behind a NAT send ICMP packets to the same nodes

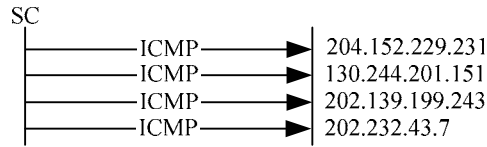

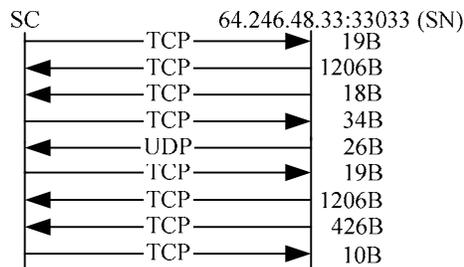

SC sends UDP packets to 4 distinct nodes and receives response over UDP. We believe that these nodes also run Skype.

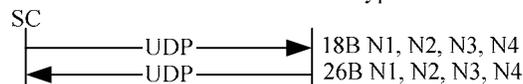

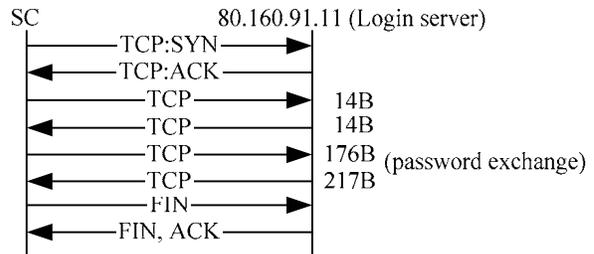

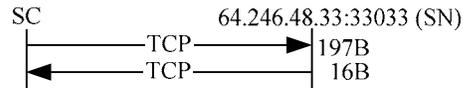

SC sends UDP packets to 22 distinct nodes and receives response from them over UDP

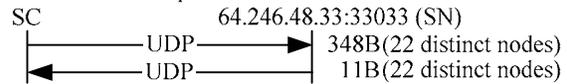

**Figure 4. Message flow for first login after installation for SC behind a simple NAT. 'B' stands for bytes and 'N' stands for node. SYN and ACK packets are shown to indicate who initiated TCP connection. Message flows are not strictly according to time. Messages have been grouped together to provide a better picture. Message size corresponds to size of TCP or UDP payload. Not all messages are shown in the message flow.**

For a SC behind a port-restricted NAT, the message flow for login was roughly the same as for a SC on a public IP address. However, more data was exchanged. On average, SC exchanged 10 kilobytes of data with SN, login server, and other Skype nodes. The message flow is shown in Figure 4.

A SC behind a port-restricted NAT and a UDP-restricted firewall was unable to receive any UDP packets from machines outside the firewall. It therefore could send and receive only TCP traffic. It had a TCP connection with a SN and the login server, and it exchanged information with them over TCP. On average, it exchanged 8.5 kilobytes of data with SN, login server, and other Skype nodes. The message flow is shown in Figure 5.

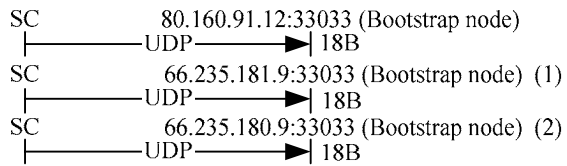

Bootstrap nodes 66.235.180.9 and 66.235.181.9 are represented by labels (1) and (2) respectively in subsequent flows

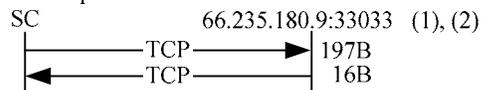

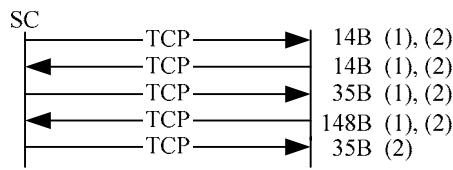

For same Skype user id, SC on public IP address and SC behind a NAT, and SC behind a UDP restricted firewall send ICMP packets to the same nodes

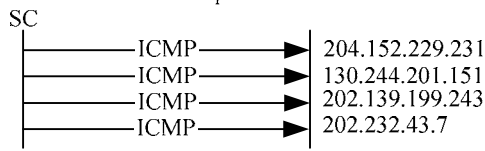

SC decides that it will retain TCP connection with 66.235.181.9. This node becomes a SN.

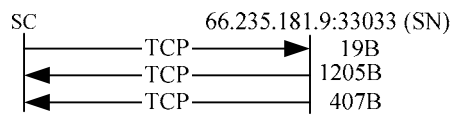

SC sends UDP packets to 4 distinct nodes. Since it is behind UDP restricted firewall, it cannot receive any responses over UDP.

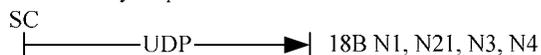

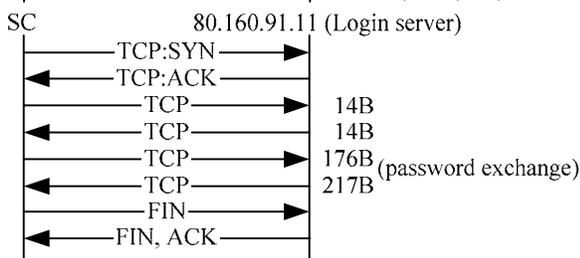

SC sends UDP packets to 18 distinct nodes. It is conjectured that they should be online Skype nodes.

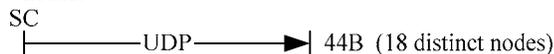

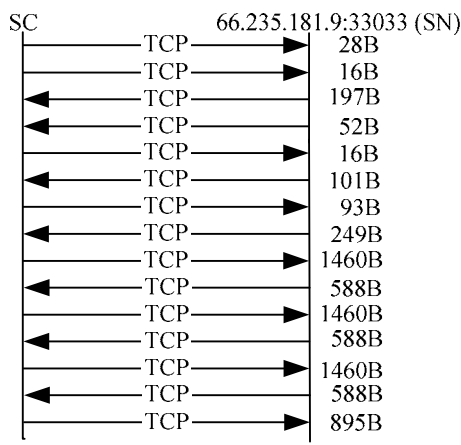

**Figure 5. Message flow for first login after installation for a SC behind a firewall, which blocks UDP packets. 'B' stands for bytes and 'N' stands for node. SYN and ACK packets are shown to indicate who initiated TCP connection. Message flows are not strictly according to time. Messages have been grouped together to provide a better picture. Message size corresponds to size of TCP or UDP payload. Not all messages are shown in the message flow.**

The following inferences can be drawn by careful observation of call flows in Fig 3, 4, and 5.

### 4.2.4.1 NAT and Firewall Determination

We conjecture that a SC is able to determine at login if it is behind a NAT and firewall. We guess that there are at least two ways in which a SC can determine this information. One possibility is that it can determine this information by exchanging messages with its SN using a variant of the STUN [1] protocol. The other possibility is that during login, a SC sends and possibly receives data from some nodes after it has made a TCP connection with the SN. We conjecture that at this point, SC uses its variation of STUN [1] protocol to determine the type of NAT or firewall it is behind. Once determined, the SC stores this information in the Windows registry. We also conjecture that SC refreshes this information periodically. We are not clear on how often a SC refreshes this information since Skype messages are encrypted.

### 4.2.4.2 Alternate Node Table

Skype is a p2p client and p2p networks are very dynamic. SC, therefore, must keep track of online nodes in the Skype network so that it can connect to one of them if its SN becomes unavailable.

From Figure 3 and 4, it can be seen that SC sends UDP packets to 22 distinct nodes at the end of login process and possibly receives a response from them if it is not behind a UDP-restricted firewall. We conjecture that SC uses those messages to advertise its arrival on the network. We also conjecture that upon receiving a response from them, SC builds a table of online nodes. We call this table alternate node table. It is with these nodes a SC can connect to, if its SN becomes unavailable. The subsequent exchange of information with some of these nodes during call establishment confirms that such a table is maintained.

It can be seen from Figure 3, 4, and 5, that SC sends ICMP messages to some nodes in the Skype network. The reason for sending these messages is not clear.

### 4.2.5 Subsequent Login Process

The subsequent login process was quite similar to the first-time login process. The SC built a HC after a user logged in for the first time after installation. The HC got periodically updated with the IP address and port number of new peers. During subsequent logins, SC used the login algorithm to determine at least one available peer out of the nodes present in the HC. It then established a TCP connection with that node. We also observed that during subsequent logins, SC did not send any ICMP packets.

### 4.2.6 Login Process Time

We measured the time to login on the Skype network for the three different network setups described in Section 3. For this experiment, the HC already contained the maximum of two hundred entries. The SC with a public IP address and the SC behind a port-restricted NAT took about 3-7 seconds to complete the login procedures. The SC behind a UDP-restricted firewall took about 34 seconds to complete the login process. For SC behind a UDP-restricted firewall, we observed that it sent UDP packets to its thirty HC entries. At that point it concluded that it is behind UDP-restricted firewall. It then tried to establish a TCP connection with the HC entries and was ultimately able to connect to a SN.

## 4.3 User Search

Skype uses its Global Index (GI) [2] technology to search for a user. Skype claims that search is distributed and is guaranteed to find a user if it exists and has logged in during the last 72 hours. Extensive testing suggests that Skype was always able to locate users who logged in using public or private IP address in the last 72 hours.

Skype is a not an open protocol and its messages are encrypted. Whereas in login we were able to form a reasonably precise opinion about different entities involved, it is not possible to do so in search, since we cannot trace the Skype messages beyond a SN. Also, we were unable to force a SC to connect to a particular SN. Nevertheless, we have observed and present search message flows for the three different network setups.

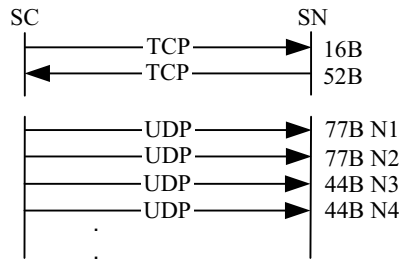

**Figure 6. Message flow for user search when SC has a public IP address. 'B' stands for bytes and 'N' stands for node. Message sizes correspond to payload size of TCP or UDP packets.**

A SC has a search dialog box. After entering the Skype user id and pressing the find button, SC starts its search for a particular user. For SC on a public IP address, SC sent a TCP packet to its SN. It appears that SN gave SC the IP address and port number of four nodes to query, since after that exchange with SN, SC sent UDP packets to four nodes. We also observed that SC had not exchanged any information with these four nodes during login. SC then sent UDP packets to those nodes. If it could not find the user, it informed the SN over TCP. It appears that the SN now asked it to contact eight different nodes, since SC then sent UDP packets to eight different nodes. This process continued until the SC found the user or it determined that the user did not exist. On average, SC contacted eight nodes. The search took three to four seconds. We are not clear on how SC terminates the search if it is unable to find a user.

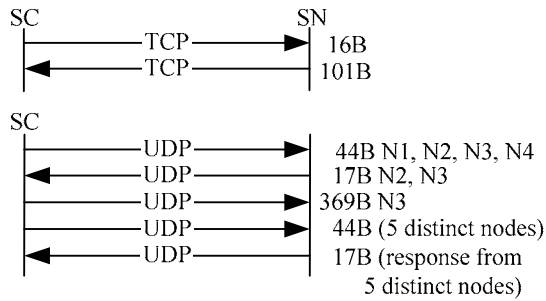

**Figure 7. Message flow for user search when SC is behind a port-restricted NAT. 'B' stands for bytes and 'N' stands for node. UDP packets were sent to N1, N2, N3, and N4 during login process and responses were received from them. Message size corresponds to payload size of TCP or UDP packets.**

A SC behind a port-restricted NAT exchanged data between SN, and some of the nodes which responded to its UDP request during login process. The message flow is shown in Figure 7.

A SC behind a port-restricted NAT and UDP-restricted firewall sent the search request over TCP to its SN. We believe that SN then performed the search query and informed SC of the search results. Unlike user search by SC on a public IP address, SC did not contact any other nodes. This suggests that SC knew that it was behind a UDP-restricted firewall. The message flow is shown Figure 8.

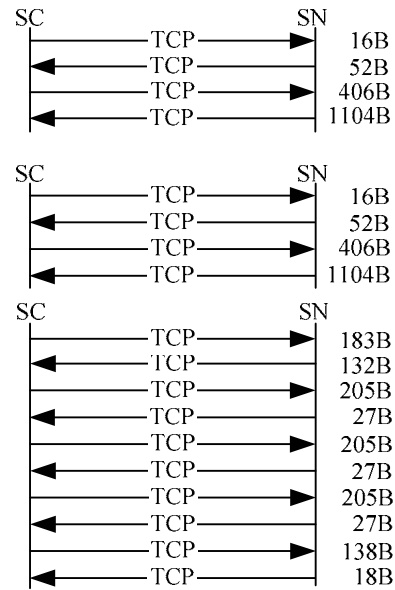

**Figure 8. User search by a SC behind a UDP-restricted firewall. 'B' stands for bytes. Data is exchanged with SN only. Message size corresponds to payload size of TCP/UDP packets.**

### 4.3.1 Search Result Caching

To observe if search results are cached at intermediate nodes, we performed the following experiment. User A was behind a port-restricted NAT and UDP-restricted firewall, and he logged on the Skype network. User B logged in using SC running on machine B, which was on public IP address. User B (on public IP) searched for user A, who is behind port-restricted NAT and UDP-restricted firewall. We observed that search took about 6-8 seconds. Next, SC on machine B was uninstalled, and Skype registry cleared so as to remove any local caches. SC was reinstalled on machine B and user B searched for user A. The search took about 3-4 seconds. This experiment was repeated four times on different days and similar results were obtained.

From the above discussion we infer that the SC performs user information caching at intermediate nodes.

## 4.4 Call Establishment and Teardown

We consider call establishment for the three network setups described in Section 3. Further, for each setup, we consider call establishment for users that are in the buddy list of caller and for users that are not present in the buddy list. It is important to note that call signaling is always carried over TCP.

For users that are not present in the buddy list, call placement is equal to user search plus call signaling. Thus, we discuss call establishment for the case where callee is in the buddy list of caller.

If both users were on public IP addresses, online and were in the buddy list of each other, then upon pressing the call button, the caller SC established a TCP connection with the callee SC. Signaling information was exchanged over TCP. The message flow between caller and callee is shown in Figure 9.

The initial exchange of messages between caller and callee indicates the existence of a challenge-response mechanism. The caller also sent some messages (not shown in Figure 9) over UDP to alternate Skype nodes, which are online Skype nodes discovered during login. For this scenario, three kilobytes of data was exchanged.

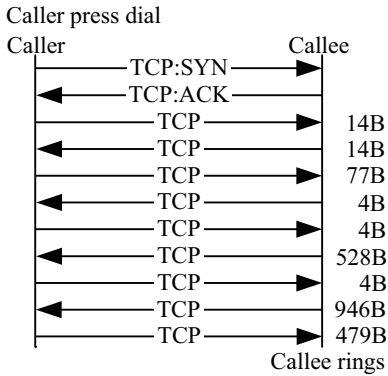

**Figure 9. Message flow for call establishment when caller and callee SC are on machines with public IP addresses and callee is present in the buddy lists of caller. 'B' stands for bytes. Not all messages are shown.**

In the second network setup, where the caller was behind port-restricted NAT and callee was on public IP address, signaling and media traffic did not flow directly between caller and callee. Instead, the caller sent signaling information over TCP to an online Skype node which forwarded it to callee over TCP. This online node also routed voice packets from caller to callee over UDP and vice versa. The message flow is shown in Figure 10.

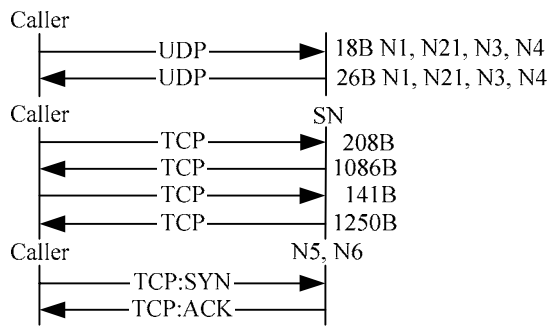

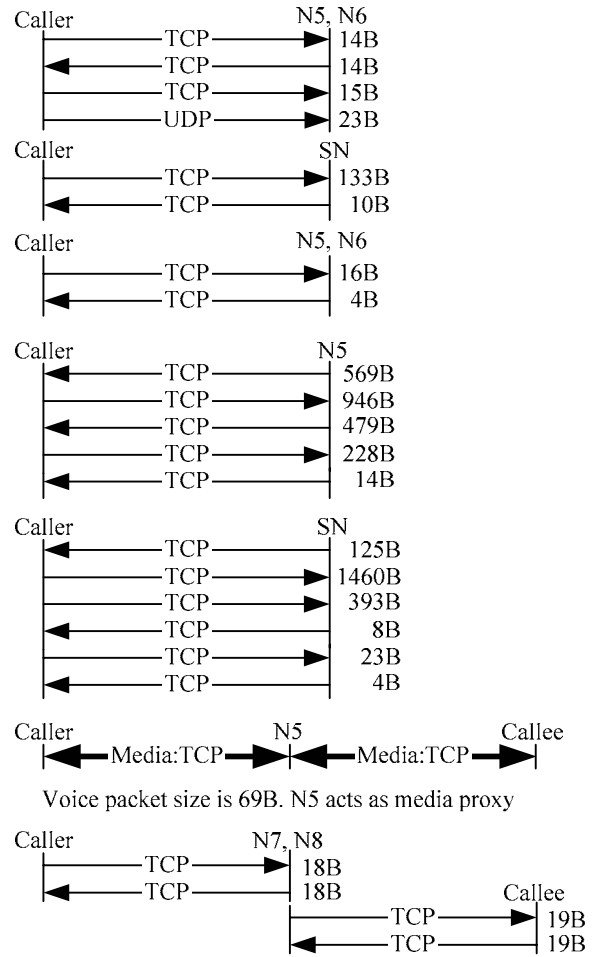

**Figure 10. Message flow for call establishment when caller SC is behind a port-restricted NAT and callee SC is on public IP address. 'B' stands for bytes and 'N' stands for node. Not all messages are shown. Caller SC sent UDP messages to nodes 5, 6, 7, and 8 during login and received responses from them. We thus believe caller SC stored the IP address and port of these nodes in its internal tables, which we call the alternate node table.**

For the third setup, in which both users were behind port-restricted NAT and UDP-restricted firewall, both caller and callee SC exchanged signaling information over TCP with another online Skype node. Caller SC sent media over TCP to an online node, which forwarded it to callee SC over TCP and vice versa. The message flow is shown in Figure 11.

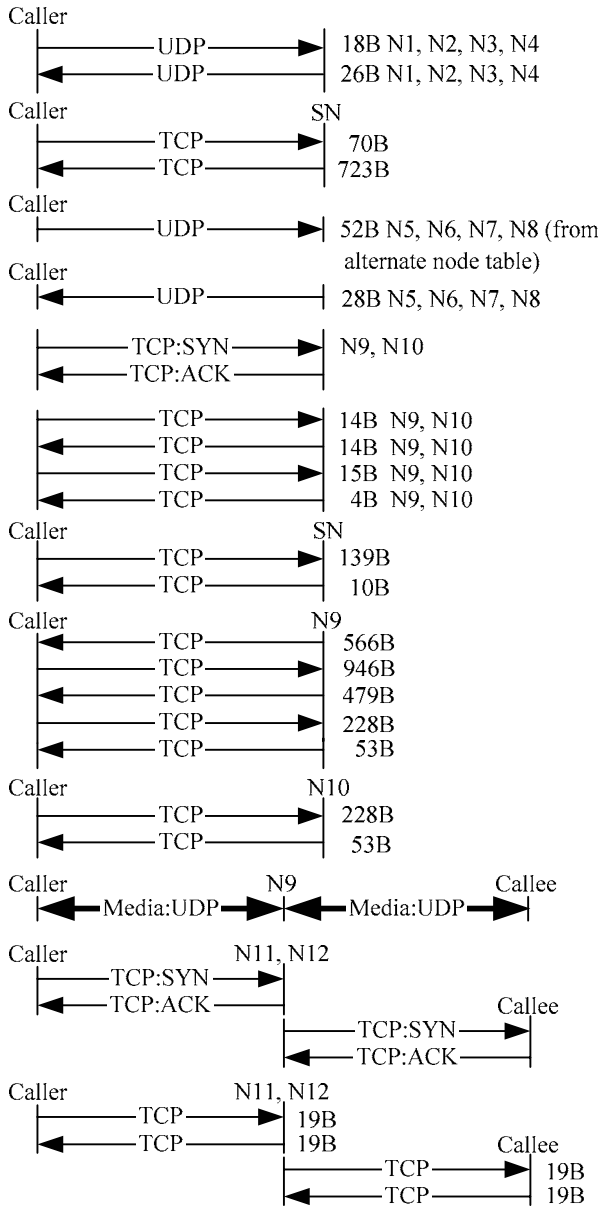

Caller and callee on the average exchange 3 msg/s over TCP with N11, and N12 during the time call is established.

**Figure 11. Message flow for call establishment when caller and callee SC are behind a port-restricted NAT and UDP-restricted firewall. 'B' stands for bytes and 'N' stands for a node. Not all messages are shown. Voice traffic flows over TCP.**

There are many advantages of having a node route the voice packets from caller to callee and vice versa. First, it provides a mechanism for users behind NAT and firewall to talk to each other. Second, if users behind NAT or firewall want to participate in a conference, and some users on public IP address also want to join the conference, this node serves as a mixer and broadcasts the conferencing traffic to the participants. The negative side is that there will be a lot of traffic flowing across this node. Also, users generally do not want that arbitrary traffic should flow across their machines.

During call tear-down, signaling information is exchanged over TCP between caller and callee if they are both on public IP addresses, or between caller, callee and their respective SNs. The messages observed for call tear down between caller and callee on public IP addresses are shown in Figure 12.

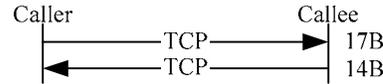

**Figure 12. Call tear down message flow for caller and callee with public IP addresses**

For the second, and third network setups, call tear down signaling is also sent over TCP. We, however, do not present these message flows, as they do not provide any interesting information.

### 4.5 Media Transfer and Codecs

If both Skype clients are on public IP address, then media traffic flowed directly between them over UDP. The media traffic flowed to and from the UDP port configured in the options dialog box. The size of voice packet was 67 bytes, which is the size of UDP payload. For two users connected to Internet over 100 Mb/s Ethernet with almost no congestion in the network, roughly 140 voice packets were exchanged both ways in one second. Thus, the total uplink and downlink bandwidth used for voice traffic is 5 kilobytes/s. This bandwidth usage corresponds with the Skype claim of 3-16 kilobytes/s.

If either caller or callee or both were behind port-restricted NAT, they sent voice traffic to another online Skype node over UDP. That node acted as a media proxy and forwarded the voice traffic from caller to callee and vice versa. The voice packet size was 67 bytes, which is the size of UDP payload. The bandwidth used was 5 kilobytes/s.

If both users were behind port-restricted NAT and UDP-restricted firewall, then caller and callee sent and received voice traffic over TCP from another online Skype node. The TCP packet payload size for voice traffic was 69 bytes. The total uplink and downlink bandwidth used for voice traffic is about 5 kilobytes/s. For media traffic, SC used TCP with retransmissions.

The Skype protocol seems to prefer the use of UDP for voice transmission as much as possible. The SC will use UDP for voice transmission if it is behind a NAT or firewall that allows UDP packets to flow across.

#### 4.5.1 Silence Suppression

No silence suppression is supported in Skype. We observed that when neither caller nor callee was speaking, voice packets still flowed between them. Transmitting these silence packets has two advantages. First, it maintains the UDP bindings at NAT and second, these packets can be used to play some background noise at the peer. In the case where media traffic flowed over TCP between caller and callee, silence packets were still sent. The purpose is to avoid the drop in TCP congestion window size, which takes some RTT to reach the maximum level again.

### 4.5.2 Putting a Call on Hold
Skype allows peers to hold a call. Since a SC can operate behind NATs, it must ensure that UDP bindings are mak at a NAT. On average, a SC sent three UDP packets per second to the call peer, SN, or the online Skype node acting as a media proxy when a call is put on hold. We also observed that in addition to UDP messages, the SC also sent periodic messages over TCP to the peer, SN, or online Skype node acting as a media proxy during a call hold.

### 4.5.3 Codec Frequency Range
We performed experiments to determine the range of frequencies Skype codecs allow to pass through. A call was established between two Skype clients. Tones of different frequencies were generated using the NCH Tone Generator [11] on the caller SC and output was observed on the callee SC and vice versa. We observed that the minimum and maximum audible frequency Skype codecs allow to pass through are 50 Hz and 8,000 Hz respectively.

Using Net Peeker [4], we reduced the uplink and downlink bandwidth available to Skype application to 1500 bytes/s, respectively. We observed that the minimum and maximum audible frequencies Skype codecs allowed to pass through remained unchanged i.e. 50 Hz and 8,000 Hz, respectively.

### 4.5.4 Congestion
We checked Skype call quality in a low bandwidth environment by using Net Peeker [4] to tune the upload and download bandwidth available for a call. We observed that uplink and downlink bandwidth of 2 kilobytes/s each was necessary for reasonable call quality. The voice was almost unintelligible at an uplink and downlink bandwidth of 1.5 kilobytes/s.

## 4.6 Keep-alive Messages
We observed in for three different network setups that the SC sent a refresh message to its SN over TCP every 60s.

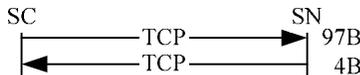

**Figure 13. Skype refresh message to SN**

## 5. CONFERENCING
We observed the Skype conferencing features for a three-user conference for the three network setups discussed in Section 3. We use the term user and machine interchangeably. Let us name the three users or machines as A, B, and C. Machine A was a 2 GHz Pentium 4 with 512 MB RAM while machine B, and C were Pentium II 300MHz with 128 MB RAM, and Pentium Pro 200 MHz with 128 MB RAM, respectively. In the first setup, the three machines had a public IP address. A call was established between A and B. Then B decided to include C in the conference. From the ethereal dump, we observed that B and C were sending their voice traffic over UDP to SC on machine A, which was acting as a mixer. It mixed its own packets with those of B and sent them to C over UDP and vice versa as shown in Figure 14. The size of the voice packet was 67 bytes, which is the size of UDP packet payload.

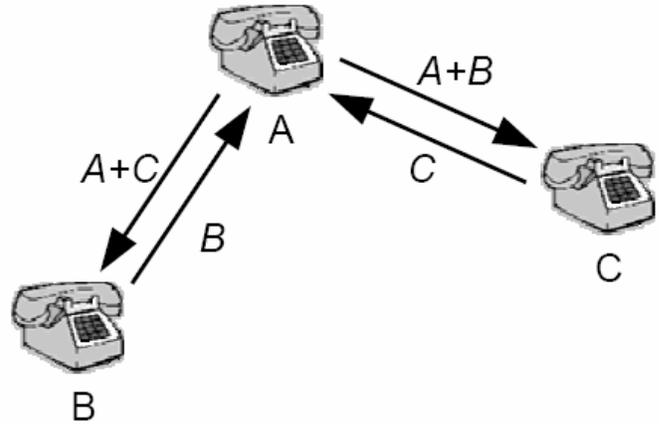

**Figure 14. Skype three user conferencing**

In the second setup, B and C were behind port-restricted NAT, and A was on public Internet. Initially, user A and B established the call. Both A and B were sending media to another Skype online node, which forwarded A's packets to B over UDP and vice versa. User A then put B on hold and established a call with C. It then started a conference with B and C. We observed that both B and C were now sending their packets to A over UDP, which mixed its own packets with those coming from B and C, and forwarded it to them appropriately.

In the third setup, B and C were behind port-restricted NAT and UDP-restricted firewall and A was on public Internet. User A started the conference with B and C. We observed that both B and C were sending their voice packets to A over TCP. A mixed its own voice packets with those coming from B and C and forwarded them to B and C appropriately.

We also observed that even if user B or C started a conference, A's machine, which was the most powerful amongst the three, always got elected as conference host and mixer.

The white paper [7] observes that if iLBC [8] codec is used, then the total call 36 kb/s for a two-way call. For three-user conference, it jumps to 54 kb/s for the machine hosting the conference.

For a three party conference, Skype does not do full mesh conferencing [12].

## 6. OTHER EXPERIMENTS
Unlike MSN Messenger, which signs out the user if that user logs in on other machine, Skype allows a user to log in from multiple machines simultaneously. The calls intended for that user are routed to all locations. Upon user picking a call at one location, the call is immediately cancelled at other locations. Similarly, instant messages for a user who is logged in at multiple machines are delivered to all the locations.

A voice call was established between a SC in the IRT Lab [20] and a SC connected to a 56 kb/s modem. Modem users were in China, Pakistan, and Singapore. The experiment was then repeated with MSN, and Yahoo messengers. In all three cases, the modem users reported better quality for Skype.

The SN is selected by the Skype protocol based on a number of factors like CPU and available bandwidth. It is not possible to arbitrarily select a SN by filling the HC with IP address of an online SC. This conclusion was drawn from the following experiment. Consider two online Skype nodes A and B. A is connected to Skype network and has only one entry in its HC. We call super node of A as SN_A. Now we modify the HC of SC on machine B, such that it only contains the IP address and port number of SC running at A. When B logged onto the Skype network, we observed that it connected to A's super node rather than connecting to A.

## 7. CONCLUSION

Skype is the first VoIP client based on peer-to-peer technology. We think that three factors are responsible for its increasing popularity. First, it provides better voice quality than MSN and Yahoo IM clients; second, it can work almost seamlessly behind NATs and firewalls; and third, it is extremely easy to install and use. We believe that Skype client uses its version of STUN [1] protocol to determine the type of NAT or firewall it is behind. The NAT and firewall traversal techniques of Skype are similar to many existing applications such as network games. It is by the random selection of sender and listener ports, the use of TCP as voice streaming protocol, and the peer-to-peer nature of the Skype network, that not only a SC traverses NATs and firewalls but it does so withhout any explicit NAT or firewall traversal server. Skype uses TCP for signaling. It uses wide band codecs and has probably licensed them from GlobalIPSound [10]. Skype communication is encrypted.

The underlying search technique that Skype uses for user search is still not clear. Our guess is that it uses a combination of hashing and periodic controlled flooding to gain information about the online Skype users.

Skype has a central login server which stores the login name and password of each user. Since Skype packets are encrypted, it is not possible to say with certainty what other information is stored on the login server. However, during our experiments we did not observe any subsequent exchange of information with the login server after a user logged onto the Skype network.

## APPENDIX

The Appendix shows the message dump of HTTP 1.1 GET requests that a SC sent to skype.com and the responses it received, when it was started by the user.

When SC was started for the first time after installation, it sent a HTTP 1.1 GET request containing the keyword installed to skype.com. This request was not sent in subsequent Skype runs. The request is shown below:

```
GET /ui/0/97/en/installed HTTP/1.1
User-Agent: Skype™ Beta 0.97
Host: ui.skype.com
Cache-Control: no-cache
```

The 200 OK response SC received for this GET request:

```
HTTP/1.1 200 OK
Date: Tue, 20 Apr 2004 04:51:39 GMT
Server: Apache/2.0.47 (Debian GNU/Linux) PHP/4.3.5
mod_ssl/2.0.47 OpenSSL/0.9.7b
X-Powered-By: PHP/4.3.5
Cache-control: no-cache, must revalidate
Pragma: no-cache
Expires: 0
Content-Length: 0
Content-Type: text/html; charset=utf-8
Content-Language: en
```

During subsequent startups, SC sent a a HTTP 1.1 GET request containing the keyword getlatestversion to skype.com:

```
GET /ui/0/97/en/getlatestversion?ver=0.97.0.6 HTTP/1.1
User-Agent: Skype™ Beta 0.97
Host: ui.skype.com
Cache-Control: no-cache
```

The 200 OK response SC received for this GET request:

```
HTTP/1.1 200 OK
Date: Tue, 20 Apr 2004 04:51:40 GMT
   Server:   Apache/2.0.47   (Debian   GNU/Linux)
   PHP/4.3.5 mod_ssl/2.0.47 OpenSSL/0.9.7b
X-Powered-By: PHP/4.3.5
Cache-control: no-cache, must revalidate
Pragma: no-cache
Expires: 0
Transfer-Encoding: chunked
Content-Type: text/html; charset=utf-8
Content-Language: en

2
96
0
```

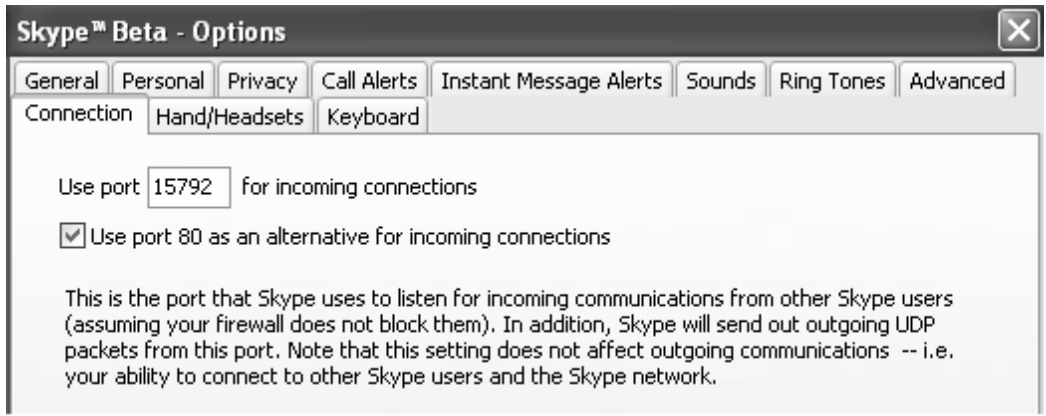

**Figure 16. Skype connection tab. It shows the ports on which Skype listens for incoming connections.**

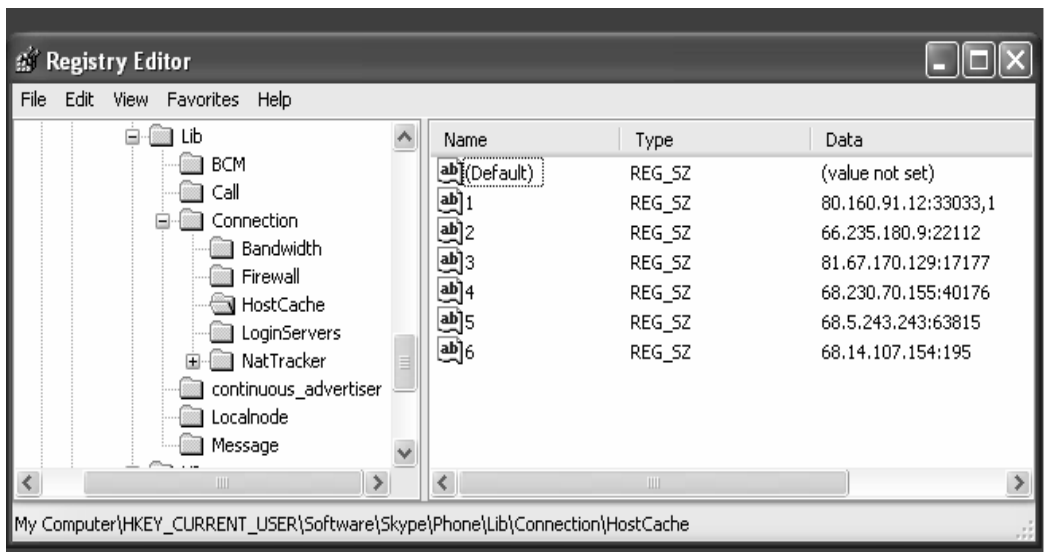

**Figure 17. Skype host cache list**